\documentstyle[12pt,hyperref]{article}
\input epsf.sty
\def\ZZZ{{\hbox{ Z\kern-1.6mm Z}}}
\def\zzz{{\hbox{ z\kern-1mm z}}}

\newcommand{\be}{\begin{equation}}
\newcommand{\ee}{\end{equation}}
\newcommand{\ben}{\begin{eqnarray}\displaystyle}
\newcommand{\een}{\end{eqnarray}}

\newcommand{\sectiono}[1]{\section{#1}\setcounter{equation}{0}}

\def\one{{\hbox{ 1\kern-.8mm l}}}
\def\zero{{\hbox{ 0\kern-1.5mm 0}}}

\begin{document}
{\baselineskip20pt
\begin{center}
{\Large \bf
String Theory and Einstein's Dream
}

\end{center} }

\vskip .6cm
\medskip

\vspace*{4.0ex}

\centerline{\large \rm  Ashoke Sen}

\vspace*{4.0ex}

\centerline{\large \it Harish-Chandra Research Institute}

\centerline{\large \it  Chhatnag Road, Jhusi,
Allahabad 211019, INDIA}

\vspace*{1.0ex}

\centerline{E-mail:  ashoke.sen@cern.ch,
sen@mri.ernet.in}

\vspace*{5.0ex}

\sectiono{Introduction} \label{s0}

Unification of the theory of gravitation, as given by 
Einstein's general theory
of relativity, and the theory of electromagnetism, as formulated by
Maxwell, had been Einstein's dream during the later part of his life.
String theory, which is the subject of this article, is an attempt to
realize this dream. However in many ways string theory
attempts to go  beyond Einstein's dream. String
theory attempts to bring all known forces of nature, -- not just
gravity and electromagnetism,  --
under one umbrella. It also tries to do
so in a manner that is consistent with the principles of quantum
mechanics, -- the theory that is necessary for describing the laws
of nature at very small distance. 
Thus string theory is an attempt to provide an
all encompassing
description of nature that works at large distances where gravity
becomes important as well as small distances where quantum mechanics
is important.

In this article I shall try to give a very general introduction to string
theory.\footnote{Refs.\cite{r1,r2,r3,r4} provide some good
introductory textbooks on string theory.}
However in order to do so, I must begin by reviewing our
current understanding of the basic constituents of matter. This is the
subject to which we shall now turn.

\sectiono{The World of Elementary Particles} \label{s1}
 
\begin{figure}
 \begin{center}
 \leavevmode
\epsfysize=8cm
\epsfbox{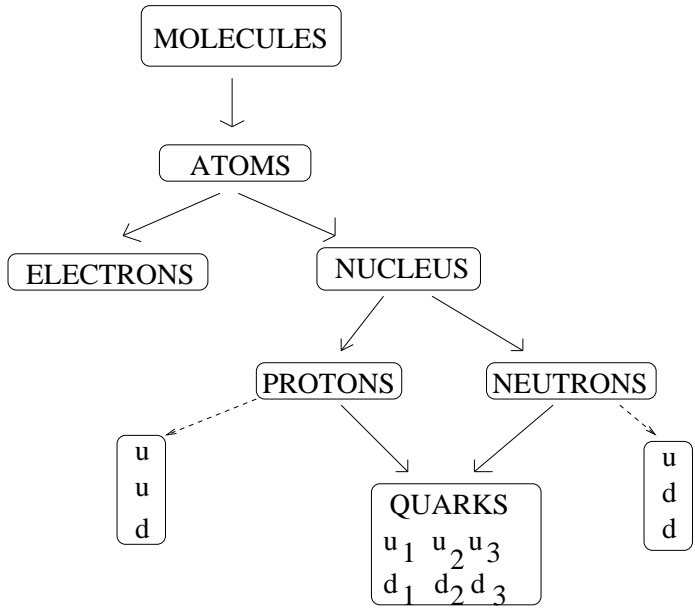}
\caption{Our current understanding of the building blocks of matter.}
\label{f1}
\end{center}
\end{figure}

According to our current understanding, everything
that we see around us
is made of a few elementary building blocks. 
Figure~\ref{f1} gives us a bird's eye view of our current knowledge
of the structure of matter.
At the crudest
level the building blocks of matter
are the individual molecules of various
compounds. However there are a very large number of 
compounds, each with 
its own characteristic molecule. A simpler picture emerged when it was
realized that each molecule is made of some smaller building blocks 
known as atoms. 
There are about 100 different types of atoms and different molecules
differ in their properties because they contain different number of atoms
of different types in different arrangements.
During the early years of the twentieth century
it was realized that atoms are also not the smallest constituents of matter,
-- each atom is made of   
a central nucleus and a set of electrons revolving around it.
Different atoms have different number of electrons, but all the
electrons found in all atoms have identical properties.
In contrast 
the nuclei of different types of
atoms have very different properties. 
This picture simplified once it was realized that each nucleus  
can be regarded as being made of even smaller constituents, -- the
proton and the neutron. Different nuclei have different 
properties because they contain different numbers of protons
and neutrons. Finally, even the protons and neutrons are now
known to be made of even smaller constituents called quarks --
the proton being made of two up ($u$) quarks and one down ($d$)
quarks, and the neutron of one $u$ and two $d$ quarks.
According to
our current knowledge, the electrons and the quarks
cannot be divided any further. We call them
elementary particles.

This gives us a very simple picture of the structure of matter,  namely
everything is made of three different types of `elementary particles' 
-- the electron, the $u$ quark and the $d$ quark. However as we shall
see, this is far from a complete picture. As is already evident from 
Fig.~\ref{f1}, the up and down quarks each come in three varieties.
Here we have denoted them by $u_1$, $u_2$, $u_3$ and $d_1$, $d_2$,
$d_3$, but often they are refered to as red, blue and green type of quarks.
We shall refer to this as the colour quantum number although this has
nothing to do with the colour that we see in everyday life.
The quarks inside the proton and neutron continuously change their
colour due to a process known as strong interaction that will be
discussed soon. There are various other reasons why this picture is not
complete. I shall review some of them here.

In order to understand the structure of matter, we need to understand
not only the basic constituents of matter, but also the nature of the forces
that operate between them. Without this knowledge we shall not have
any understanding of what keeps the quarks bound inside a proton and
neutron, or at a larger scale, of what keeps the atoms bound inside a 
molecule. According to our current knowledge there are four basic
types of forces operating between elementary particles, -- 1) gravitational,
2) electromagnetic, 3) strong and 4) weak. Of these the gravitational and
the electromagnetic forces are familiar to us from everyday experience.
For example the gravitational force is responsible for earth's gravity and
the motion of the planets around the sun. The electromagnetic force is
the cause of lightening in the sky,  
the force of a magnet, working of various electrical
appliances etc. It is also
responsible for binding the electrons and the  nuclei inside the atom
and the atoms inside a molecule.
The strong force operates between quarks and is responsible for 
binding them inside a proton and a neutron and also for binding
the proton and the neutron inside a nucleus. The weak force, being weak,
is not responsible for binding any particles; however it is responsible
for certain radioactive decays known as $\beta$-decay.

It turns out that in studying the physics of elementary
particles, we can ignore the effect of gravitational force.
To see this one can compare the electrostatic force between two
protons with the gravitational force between two protons at rest.
The result is
$$ {\hbox{Grav. Force}\over \hbox{Elec. Force}} = {G_N
m_p^2/r^2\over
e_p^2/r^2}
\sim 10^{-36} $$
where
$G_N$ is the Newton's constant
($6.67 \times 10^{-8}$ cm$^3$/gm sec$^2$) that controls the
strength of the gravitational force between two bodies,
$m_p$ is the proton mass ($1.67 \times 10^{-24}$ gm)
and 
$e_p$ is the proton charge ($4.8 \times 10^{-10}$ e.s.u.).
Clearly this ratio is extremely small.
Similarly all other forces can also be shown to be much larger than the
gravitational force.  

\begin{figure}
\begin{center}
\leavevmode
\epsfysize 5cm
\epsfbox{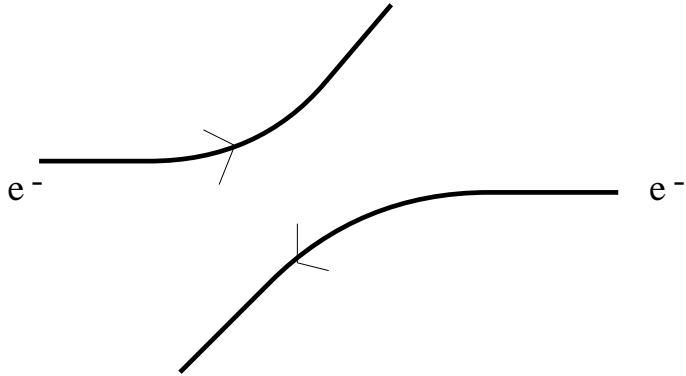}
\caption{Classical picture of the 
deflection of a pair of electrons via electromagnetic
force.} \label{f2}
\end{center}
\end{figure}

\begin{figure}
\begin{center}
\leavevmode
\epsfysize 5cm
\epsfbox{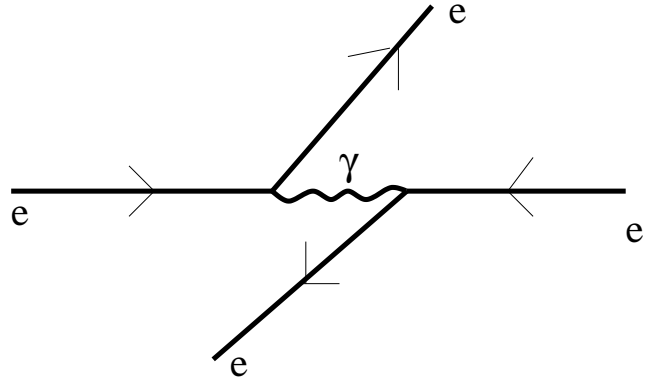}
\caption{Quantum picture of the
deflection of a pair of electrons via electromagnetic
force.} \label{f3}
\end{center}
\end{figure}

So far we have discussed the elementary particles and the forces
operating between them as separate entities, but with
 the help of quantum theory one can
give a unified description of elementary particles, and the
forces among the elementary particles.
Consider for example the electromagnetic force between two
electrons when they pass each other.
Due to this force, each particle gets deflected from its
original trajectory. This has been depicted in Fig.~\ref{f2}.
In quantum theory, one provides a different explanation of the
same phenomenon. Here
the deflection takes place because the
two electrons exchange a new particle, called photon, while
passing near each other (see Fig.~\ref{f3}).
The photon is capable of carrying some amount of energy and
momentum from the first electron to the second electron, thereby
causing this deflection.\footnote{The quantum picture shown 
in Fig.~\ref{f3} suggests that the change
in the direction of the electrons happens suddenly instead of
continuously. In practice each exchange of photon causes a tiny amount
of sudden jump, and the classical picture emerges due to the
quantum process repeating many times 
via many exchanges of photons.}
We call the photon the mediator of electromagnetic 
force. Even though it mediates electromagnetic force, the photon
itself is electrically neutral.

Thus in the language of quantum theory 
we can describe a force by specifying
the particle(s) which mediate the force.
It turns out that
the strong force is mediated by eight different particles known
as gluons.
These particles are all electrically neutral. The quarks inside a
proton (and neutron) continuously exchange gluons, and in this
process keep changing their colour quantum number.
On the other hand the weak force is mediated 
by three particles, denoted by
$W^+$, $W^-$ and $Z$.
$W^+$ and $W^-$ carry +1 and -1 unit of electric 
charge respectively while
$Z$ is neutral. (The unit of electric charge is taken to be the charge carried
by a single proton. Thus $W^+$ has charge equal to that of a proton,
while $W^-$ has charge that is equal in magnitude but opposite
in sign to that of a proton.)

Clearly, we must add the gluons, $W^+$, $W^-$ and $Z$,
as well as the photon, to our list
of elementary particles. We shall refer to these as the mediator
particles.  
Theoretical analysis shows that for every elementary particle there
must also be another elementary particle, known as the antiparticle,
that carries exactly the same amount of  charge but with 
opposite sign. Thus for every quark and the electron we have the
corresponding anti-quark and the anti-electron (known as the
positron). Fortunately the gluons, the photon and the $Z$ particles
are their own anti-particles, whereas $W^-$ is the anti-particle of
$W^+$ and vice-versa. Thus we do not need to expand our list by
including anti-particles of the mediator particles.
However this still does not exhaust the list of all
elementary particles.
Besides the $u$ and $d$
quarks, electrons and mediators and their anti-particles, 
there are also other elementary 
particles which are produced by cosmic rays, radioactive decays, collision 
of high energy particles, etc.
They must also be added to the list.

Our current list contains about 100 types of
elementary particles. Thus the situation
would not seem any better than the days when atoms were thought
to be the basic constituents of matter. The properties
of matter known at that time could be explained 
in terms of the properties of
about 100 types of atoms. There is however a difference,
-- unlike the case of atoms, there is a simple mathematical theory
that explains the properties of all
the elementary particles. In fact this
theory has been so successful that it has come to be known as the
`standard model' of elementary particles.
This model, in principle, can be used to calculate the
result of any experiment that we wish to perform involving the
elementary particles.
So far the standard model 
has been extremely successful in
explaining almost all experimental results.

\sectiono{The Standard Model: Its 
Successes and Limitations} \label{s2}

In this section I shall explain some of the basic properties
of the standard model. The basic inputs in this theory are
\begin{itemize}
\item quantum mechanics, 
\item special theory of relativity, and
\item laws of electromagnetism and their generalization to 
strong and weak forces.
\end{itemize}
There is a mathematical framework, known as gauge theory, 
that includes all these three
features. I shall not describe the details of this
framework here. It turns out that there are many different
consistent gauge theories, one of which 
describes the theory of elementary particles. This particular 
theory
is known as the standard model.

Once the theory is written down, it predicts the outcome of
every possible experiment involving elementary particles. (Of
course some experimental inputs go in to decide on what is the
right theory.) For example the standard model tells us precisely
what kind of elementary particles we have in our world. 
According to this model, the elementary particles in our world
fall into four categories:
 
\begin{itemize}
{\bf \item Quarks} \qquad
$ u_1, u_2, u_3$, \quad $d_1, d_2, d_3$, \quad $ c_1, c_2, c_3$,
\quad $s_1, s_2, s_3$, \quad $t_1, t_2, t_3$, \quad $b_1, b_2, b_3$

\noindent In this list we recognize the familiar up 
and down
quarks, each coming in three colours. It turns out that nature
contains four more types of quarks, -- charm ($c$), strange ($s$),
top ($t$) and bottom ($b$), each coming in three colours. These
four types of quarks are not usually found inside matter but can
be produced in highly energetic collision among normal matter.
Of the six quarks, the up, charm and top quarks carry 2/3 unit of
electric charge, whereas the down, strange and bottom quarks
carry $-1/3$ unit of electric charge.
For each quark we also have its anti-quark; we have not listed them
separately here.

 {\bf \item Leptons} \qquad 
$e^-, \nu_e$, \quad $\mu^-, \nu_\mu$, \quad $\tau^-, \nu_\tau$

\noindent In this list we recognize the electron ($e^-$); the $-$ sign
on top is to remind ourselves that the electron carries $-1$
 unit of charge, \i.e. charge equal in magnitude but opposite in sign
 to that carried by the proton. $\nu_e$, -- known as the electron
 neutrino, -- is a weakly interacting chargeless particle. These are
 so weakly interacting that a neutrino passing through the 
 earth does so
 experiencing almost no force. The pair
 of particles$(\mu^-, \nu_\mu)$ have
 properties similar to that of the pair
 $(e^-, \nu_e)$ although the muon
 ($\mu^-$) is a lot heavier that the electron. Similarly the pair
$(\tau^-, \nu_\tau)$ have properties similar to that of 
$(e^-, \nu_e)$, with the tau particle $(\tau^-$) being even heavier
than a muon. For each lepton we also have an anti-lepton which
we have not listed here. For example, the anti-particle of 
the electron is called the positron and denoted by the symbol $e^+$.

{\bf \item Gauge Bosons}\qquad 
gluons:   $g_1, \ldots g_8$,  \quad Photon:  $\gamma$,
\quad
$W^+$, $W^-$,   $Z$

\noindent These are the by now familiar mediator particles which have
been discussed before. As already mentioned the list is complete
without having to add the anti-particles separately.

{\bf \item Higgs Particle} \quad $\phi$

This is the most mysterious particle in the standard model. Unlike
every other particle in the list which has been experimentally
observed, the Higgs particle has never been seen in any experiment
despite several attempts. Nevertheless its existence is predicted by the
standard model, and new experiments are being designed to look for
this particle.

\end{itemize}

\begin{figure}
\begin{center}
\leavevmode
\epsfysize 5cm
\epsfbox{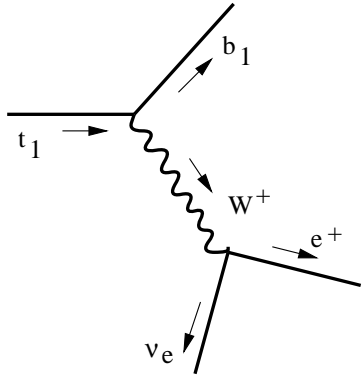}
\caption{An allowed process in the standard model.} 
\label{f4}
\end{center}
\end{figure}

The standard model not only gives us a list of elementary particles
but also the list of processes that can occur involving these particles.
For example in order to explain the electromagnetic force between
electrons using the process described in Fig.~\ref{f3}, it is
necessary to know that an electron can emit a photon. This follows
from the mathematical framework that lies behind the standard
model. The same mathematical framework also tells us that
if in this diagram we replace the electron by an electron neutrino
then this is not an allowed process in the standard model; hence a
neutrino cannot exchange a photon with another particle.
Fig.~\ref{f4} shows another example of a process that can occur
in the standard model. This describes the decay of a top quark 
($t_1$) into
an electron neutrino ($\nu_e$), a positron ($e^+$) and a bottom
quark ($b_1$).
In fact
the standard model not only tells us which processes can occur, but
it also gives us precise mathematical formula for calculating the
probablility of occurance of any such process. These predictions
are then compared with experimental data to test the model.

Given the success of the standard model, one might like to
conclude that we now have a complete understanding of the
elementary constituents of matter. This however is not true.
There are several reasons why standard model cannot be the
complete theory of elementary particles. I shall review a few of
these here.
 
 First and foremost, the standard model
does not explain the origin
of one of the important forces that we observe in
nature, namely the gravitational force. In particular the list of
particles predicted by the standard model does not contain any
particle that mediates gravitational force. The effect of this
omission of course is not seen in any of the experiments involving
elementary particles since, as observed earlier in this article, the
gravitational force between two elementary particles is extremely
small compared to the other forces. Nevertheless a complete theory
must account for every possible tiny effect that exists in nature. Thus
a theory that does not provide an explanation of the gravitational
force cannot be a complete theory of nature.

In order to appreciate the gravity of this problem, 
let us first take stock of what is
known about gravity. Our current theoretical 
understanding of the gravitational force is based on the `general theory
of relativity', -- a theory written down by Einstein almost a hundred
years ago. This theory has been enormously successful in explaining
all effects related to gravity. Unfortunately this theory is based on the
principles of  classical mechanics and not of
quantum mechanics. Since other forces in nature
follow the rules of quantum mechanics, any theory that attempts to
explain gravity as well as the other forces of nature must treat
gravity according to the rules of quantum mechanics. Hence
the general theory of relativity, despite being so successful, cannot be
the final story about gravity. In fact the 
reason that this theory has been so successful so
far is that for gravity the difference between the predictions 
of a classical and the quantum theory is extremely tiny and cannot
be observed in any of the current experiments. 
(We say
that quantum effects involving gravity are extremely small.) 

Thus the problem at this stage seems to be to first find a quantum 
theory of gravity and then combine this with the standard model to
arrive at a complete theory of all elementary particles and forces
operating between them. At the first sight the problem does not
seem unsurmountable. After all, we normally obtain a quantum theory
by first writing down a classical theory and then applying a definite
set of rules to turn it into a quantum theory. Why can't the same thing
be done with the general theory of relativity? If one  proceeds
to do this one does get some encouraging results at first. 
In particular one finds
that like other forces, gravity is also mediated by a new
kind of elementary particle. This particle has been given the name
graviton. Like the diagram in Fig.~\ref{f3} one will have a diagram 
where two electrons exchange a graviton, representing
the (tiny amount of) deflection of one of the electrons
due to gravitational force of the other electron.

\begin{figure}
\begin{center}
\leavevmode
\epsfysize 5cm
\epsfbox{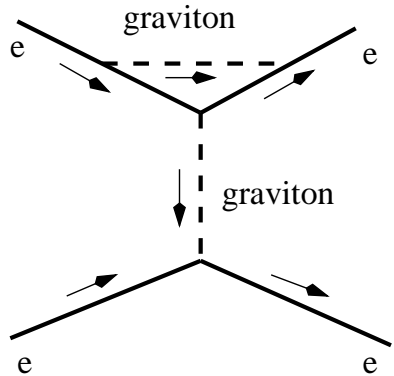}
\caption{An infinite contribution to the gravitational scattering
of two electrons.} 
\label{f5}
\end{center}
\end{figure}

So far everthing seems to be proceeding as desired. 
However one soon runs into a problem with this 
approach. To
understand the origin of this difficulty consider the process shown in
Fig.~\ref{f5} involving multiple 
graviton exchanges.  As in the case of the standard model, 
there are precise
mathematical rules for computing the probability amplitude of this
process in the quantum general theory of relativity. When one applies
those rules to calculate this probability amplitude, one finds that the
result is infinity!

This is clearly a nonsensical answer! In actual practice we know that
this probability must be extremely tiny since no experiment has yet seen
the effect of gravitational force between elementary particles. Thus there
must be something wrong with this theory.

In order to appreciate how string theory eventually resolves this problem,
it will be useful to investigate in a little more detail the origin of this
problem. You would notice that in a diagram like the one shown in
Fig.~\ref{f5} there are `interaction vertices' where three (or more)
lines meet. 
For example in Fig.~\ref{f5} there are four such interaction vertices.
These are the points where something happens.
We can regard these points as the basic events which make up the
complete process. Each such event takes place at a
given point in space at a given time, and in order to calculate the total
probability amplitude of the process we must integrate over the location
of each event in space as well as in time.  It turns out that the integrand,
calculated using the rules of quantum theory, diverges (becomes
infinite) when more than two or more such elementary events take
place at the same point in space at the same time. This in turn causes 
the integral to diverge occasionally.\footnote{Similar divergences
also occur in the standard model, but can be removed by a procedure
known as renormalization. This procedure does not work for
general theory of relativity since the divergences are more severe.}

In any case the final outcome of this complicated analysis is that the
standard procedure that has been successful in formulating a quantum
theory of strong, weak and electromagnetic forces do not work for 
gravity, and for this reason it is not easy to incorporate gravity 
into the standard model.

Besides the problem of incorporating gravity, 
the standard model suffers
from other conceptual and technical problems. While it is
true that the standard model, once formulated, can predict the results
of most experiments involving elementary particles, the formulation
of the theory itself requires a lot of input from experiments. For example
there are many consistent gauge theories, often labelled by several
continuous parameters, and standard model corresponds to one of these
theories with a specific choice of the values of these parameters. There
is no explanation within the theory as to why this particular gauge theory
with this particular choice of parameters should describe our universe.
Furthermore the choice of parameters which describes the standard
model are not generic, but 
requires  very fine tuning. This is evident from the fact that
the theory has some 
extremely small dimensionless numbers like the ratio of gravitational
and electromagnetic force between two elementary particles. For a generic
choice of parameters this ratio would be of order one. Finally recent
experiments show that not all predictions of the standard model are
completely correct. In particular, according to the standard model the
neutrinos are zero mass particles, but recent experiments show that
neutrinos actually have a tiny but finite mass. This requires a small
modification of the gauge theory that describes the standard model.

These are some of the reasons why we believe that the standard model
is not the final story. In the rest of this article we shall try to see how
string theory attempts to address some of these issues.

\sectiono{String Theory} \label{s3}

\begin{figure}
\begin{center}
\leavevmode
\epsfysize 5cm
\epsfbox{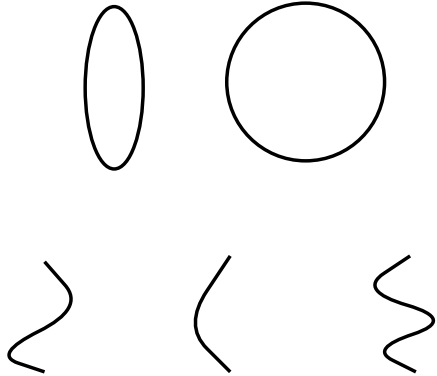}
\caption{Vibrating closed and open strings.} 
\label{f6}
\end{center}
\end{figure}

The basic idea in string theory is quite simple. It says that the
elementary constituents of matter are not point like
objects (particles) but one dimensional objects. These one
dimensional objects, also known as the fundamental (or elementary)
strings, have very specific properties which determine the various
modes in which the string can vibrate. However to the present day
experimentalists these strings appear as particles since their
size is small compared to the distance scale that can be probed by
the most powerful microscopes available today.\footnote{The most
powerful microscopes available today are in fact the particle 
accelerators. In these machines we accelerate particles to a velocity
close to that of light so that they carry very high energy and then collide
them with other particles. This process has the capability of (indirectly)
probing the structure of matter to a very small scale.
The minimum distance that can be resolved by the current accelerators
is about $10^{-16}$cm.
}
In particular,
different vibrational states of a fundamental string appear to us
as different elementary `particles'  
just as the different
modes of vibration of a single musical string can produce different
harmonics of a note.  Thus in string theory instead of having different 
types of elementary particles we have one single type of elementary
string as the basic constituent of matter. Fig.~\ref{f6} shows some of
the vibrational states of strings. As is evident from this figure, strings
can come in two varieties, -- closed strings which have no boundary
and open strings which have two end points forming its two
boundaries.

Since quantum mechanics and special theory of relativity are two
of the basic inputs in the standard model, and since
string theory must include the standard model if it is to describe
our universe, it is natural to require that string theory
also respects the principles of
quantum mechanics and special theory of relativity. 
However one finds
that for various technical reasons
it is not easy to respect these principles. 
In fact the only way we can
respect these principles is by formulating 
string theory not in the usual
three dimensional space but in a hypothetical nine dimensional 
space.\footnote{We often count time as an additional dimension
and describe this as a ten dimensional space-time. But in this article
we shall only count the number of space dimensions.}
Furthermore in this nine dimensional space one can formulate
altogether five different types of string theory, -- known as the
Type I, Type IIA, Type IIB,
E$_8\times$E$_8$ heterotic and  SO(32) heterotic string theories.
These five string theories differ from each other in the type of
vibrations which the string performs. As a result they 
have different vibrational states, which is reflected in the spectrum
of elementary `particles' that each of these theories produce.

Having nine space dimensions instead of three 
seems to be a serious problem. 
We shall return to this issue shortly and show that this in fact is not
a very serious problem. 
However, let us leave aside this problem for a moment
and discuss some of the good things
which string theory provides.
First of all one finds that one of the vibrational states of string theory have
properties identical to that of a
graviton, -- the mediator of gravitational force.
Furthermore one finds that
string theory calculations do not suffer from any
infinities of the type we encounter while trying to directly quantize
general theory of relativity. Thus string theory provides us with a finite
quantum theory of gravity!

\begin{figure}
\begin{center}
\leavevmode
\epsfysize 5cm
\epsfbox{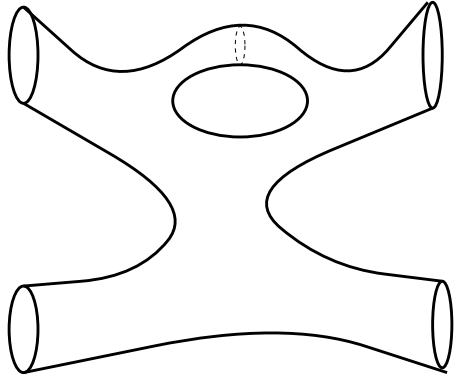}
\caption{A process describing a pair of strings scattering from each
other.} 
\label{f7a}
\end{center}
\end{figure}

It is instructive to try to understand why the probability
amplitudes calculated in string theory are finite. For this we need
to look at the figure \ref{f7a} describing the process of scattering of
two strings. Like in the case of  point particle theories,  there are
definite mathematical rules for calculating 
the probability amplitude of
this process. The point to note is that in this diagram there
are no points where specific
events (like splitiing of a single string into a
pair of strings) take place; the diagram is completely smooth 
everywhere. As a result the divergences in the point particle theories,
-- which arise when two or more such events take place at the same
point at the same time,  -- are absent in string theory. This is the
intuitive reason why string amplitudes are finite.

At this point we must mention that the graviton is only one of
the many vibrational states of an elementary string.
In fact the laws of quantum mechanics
tells us that a single elementary string has infinite number of
vibrational states. Since each such vibrational state behaves as
a particular type of elementary particle, string theory seems to 
contain infinite types of elementary particles. This would be in
contradiction with what we observe in nature were it not for the
fact that most of these elementary particles in string theory turn
out to be very heavy, and not observable in present experiments.
Thus there is no immediate conflict between what string theory
predicts and what we observe in actual experiments. On the other
hand these additional heavy elementary particles are absolutely
essential for getting finite answers in string theory. 

\begin{figure}
\begin{center}
\leavevmode
\epsfysize 5cm
\epsfbox{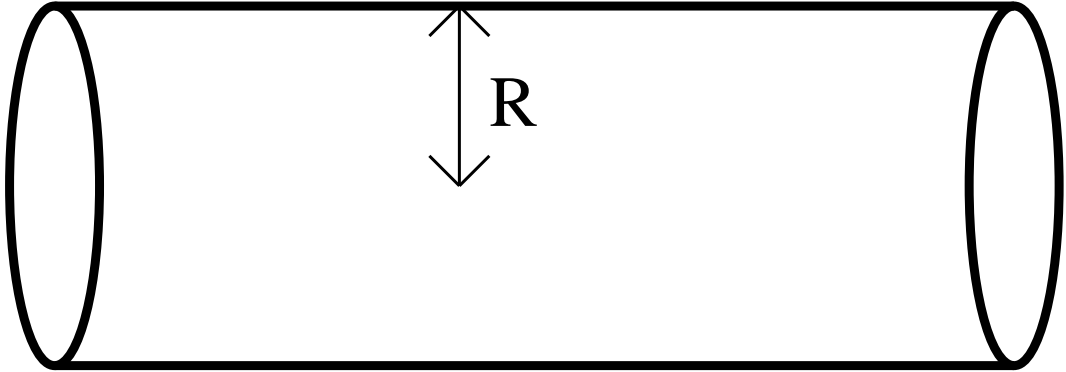}
\caption{A two dimensional space with a compact coordinate.} 
\label{f7}
\end{center}
\end{figure}

Let us now return to the issue about the dimension of space-time.
Consistency of string theory demands that we can formulate the theory
only in 9 space dimensions.
How can string theory be relevant for describing nature, which seems to
have only 3 space dimension?
The answer to this question is provided by an old idea known as
compactification. This idea
was pioneered by Kaluza and Klein during the first half of the
twentieth century and Einstein himself had been attracted by this
idea. We shall illustrate the basic idea by a simple example in which
we begin with a world with two space dimensions instead of nine
space dimensions.
We take the two space coordinates to describe the surface of a
cylinder of radius R instead of an infinite plane as shown 
in Fig.~\ref{f7}. All objects (including light) in this
world can move only along the surface of the cylinder.
Thus if we move along the vertical direction in the figure, 
then after
travelling a certain distance ($2\pi R$ where $R$ is the radius of the
cylinder) we shall 
traverse the whole circumference of the circle and
come back to the original point where we started. We call this a
compact drection. In contrast an object can travel 
along the horizontal direction
without ever returning to its original position and we call this the
non-compact direction.

\begin{figure}
\begin{center}
\leavevmode
\epsfysize .3cm
\epsfbox{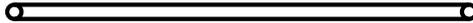}
\caption{A two dimensional space with a 
small compact coordinate.} 
\label{f8}
\end{center}
\end{figure}

Clearly if $R$ is very large (larger than the range of the most powerful
telescope) then the two dimensional space will appear to be infinite in
both directions and we would not know that one of the directions is
compact. 
If $R$ is within the visible range, then the two
dimensional creatures will start seeing infinite number 
of images of each
object separated by an interval of $2\pi R$ since light from any
object can reach an observer in  many (infinite number of) ways,
-- directly, travelling once around the circumference, travelling twice
around the circumference etc.  This may seem strange from our
point of view but will not at all seem strange from the point of view
of the two dimensional people living in this world since they would 
always see their world this way. But now consider the case when 
$R$ is very small, as shown in Fig.~\ref{f8}. 
Clearly this world looks one dimensional as $R\to$0.
In fact as 
long as $R$ is smaller than the resolution of the most
powerful microscope, the two dimensional people will never
know that they have a hidden dimension in their world. To them
the world will appear to be one dimensional.

This illustrates the way a universe with a certain number
of space dimensions can `appear to be' a universe with less
number of dimensions. This idea can be generalized to make the
nine 
dimensional world of string theory look like
three dimensional world in which we live. All we need to do is to
take  six of the nine space  directions to be small,
describing a compact space $K$.
When the size of K is sufficiently small, the space will appear to
be 3 dimensional. The main difference with the two dimensional
example that we
discussed is that while there is only one one dimensional
space (namely the circle) that can be used for making one direction
compact, there are more possibilities in higher dimensions.
An important class of six dimensional spaces which are useful for
compactification of string theory are the so called Calabi-Yau
spaces. There are many different six dimensional Calabi-Yau
spaces, and the theory that describes
the  three dimensional  world after compactification depends
on the choice of the compact space $K$, as well as which of the
five string theories we start from in nine dimensions.

Often the three dimensional theory found this way comes very close
to describing the world we see around us. In particular when we 
examine the vibrational states of the string in such a space, not only
do we find the graviton, but we often find `gauge bosons', -- the
kind of particles which mediate strong, weak and electromagnetic
forces.  Some other vibrational states have properties similar
to those of various quarks, leptons, Higgs particle etc.
Thus string theory has the potential of describing a unified
theory of elementary particles and
all the forces operating between them.

We would like to emphasize here that in string theory we use
quantum mechanics and special theory of relativity as basic
inputs; but the 
general theory of relativity and gauge theories come out of string
 theory. Thus string theory in a sense provides an explanation of why
the forces operating in our universe are described by general theory
of relativity and gauge theories. 

Of course all is not well at this stage. First of all, 
we have the
problem that even though we know of many string compactifications
which come very close to describing the world that we see, there is no
known compactification that describes exactly the world that we see
around us. Trying to look for a string compactification that describes
exactly the theory that governs our universe is an active area of
research in which many theorists are participating.
Second,
one might wonder what
is the basic principle that one uses to decide which of the five string
theories is the right theory for describing our universe. If we are
looking for a theory that describes everything in our universe,
wouldn't it be nicer to have a single mathematically
consistent theory rather than five consistent theories? Finally, even 
if there is some principle that tells us which of the five
string
theories we should use, there are still many different choices of the
compact space that brings us down to three dimension; and one might
wonder what principle decides on the choice of the compact space.
In fact it is possible to have string compactification
where the number of
non-compact direction is different from three; 
all it requires to have
$d$ non-compact directions is to choose
an appropriate compact space of dimension $(9-d)$. 
Thus the question
arises as to why our world is three dimensional?  
We shall try to address some of these issues in the next
section.

\sectiono{Duality, M-theory and the Early Universe} \label{s4}

So far we have discussed the role played by the
vibrational states of a single
fundamental string. However these are not
the only possible objects in string theory. String theory contains many
other types of objects which can be made of more than one 
(some time infinite number of) fundamental strings. We shall call these
objects composite objects.

In conventional approach to the study of elementary
constituents of matter,  we make a clear distinction between 
elementary and composite objects. For example in the standard
model the quarks are elementary particles while the proton and the
neutron are composite particles made of quarks. The standard
model tells us various properties of quarks and other elementary
particles in the theory; the properties of protons, neutrons and other
composite objects can be derived from the properties of these
constituent particles. Thus elementary particles enjoy a 
previlaged
position in the description of the theory.

The initial formulation of string theory was based on the same
principle, with the role of elementary particles being
taken over by the
elementary strings. The vibrational states of the elementary string
were the analogs of the elementary particles; all other objects made
of more than one elementary strings were composite objects whose
properties could in principle be derived from the properties of the
elementary string. However this picture, that gives a special
role to the elementary particles, got modified dramatically
after the discovery of duality symmetries in string theory. This is
the story to which we now turn.

During the mid 90's it was realised that some time a 
pair of theories which `look' different may actually describe the same
physical theory.
In other words, the same physical theory may have 
different descriptions
as different compactifications of different string theories.
This symmetry, relating the two apparently different theories,
is known as the duality symmetry. This name is actually a
misnomer, since often one finds more than two descriptions
of the same physical theory.
One of the surprising features of  duality symmetries is that
a particle which looks elementary in one description may 
appear as composite in a dual description. Thus whether a 
given particle
is elementary or composite is not an
intrinsic 
property of the particle, but
depends on which particular 
description we use for the string theory under study.

Another aspect of duality is that typically the coupling constant
of the  theory, -- the parameter that determines the strength of various
forces operating between the elementary particles  -- is
related to the coupling constant of the dual theory in a complicated
way. Due to this one finds that often a weakly coupled theory, 
\i.e. a theory with small value
of the coupling constant is related by duality to a theory with large
value of the coupling constant. Since it is easier to do calculations in
a theory for small value of the coupling constant, often duality
relates the results of a complicated calculation in one theory to the
results of a simple calculation in the dual theory.\footnote{Due to
the difficulty in doing calculations in a strongly coupled theory, most
of the dualities have not been proven, but have been tested in many
different ways.}

It is best to illustrate this with some examples. We begin with an
example of duality involving theories with all nine
dimensions non-compact. We had earlier introduced
five different consistent string theories in nine dimensions. It turns
out that the type I string theory and the $SO(32)$ heterotic string
theory are dual to each other in the sense described above. They `look'
different because the set of elementary particles, obtained from the
states of the elementary string, are quite different in the two theories.
However when one considers the full set of particles -- elementary 
and composite -- in the two theories, one finds that the 
two sets are identical. The coupling constant of the heterotic string
theory turns out to be equal to the inverse of the coupling constant
of the type I theory. Thus when the heterotic string is weakly
coupled the type I string is strongly coupled and vice versa.

Another example of duality involves string theories with five 
non-compact space directions. We take any one of the two
heterotic string  theories and take four of the space directions
to be compact, each describing a circle of certain radius. Such
a four dimensional space is known as a four torus, denoted
be the symbol $T^4$. On the other side we take type IIA string
theory and make four of the space directions compact, this time
describing a more complicated four dimensional space known as
K3. It turns out that these two five dimensional string theories 
are dual to each other.

In special cases a particular compactification of string theory may be
related to itself by a duality symmetry. In this case the duality
symmetry will relate the elementary and composite particles in the
same theory. Such theories are known called self-dual. For example
type IIB string theory with all directions non-compact
is a self-dual theory. Another
example is any of the two heterotic string theories with six compact
directions, each described by a circle. In both these theories duality
typically relates an elementary particle to a composite particle.

\begin{figure}
\begin{center}
\leavevmode
\epsfysize 7cm
\epsfbox{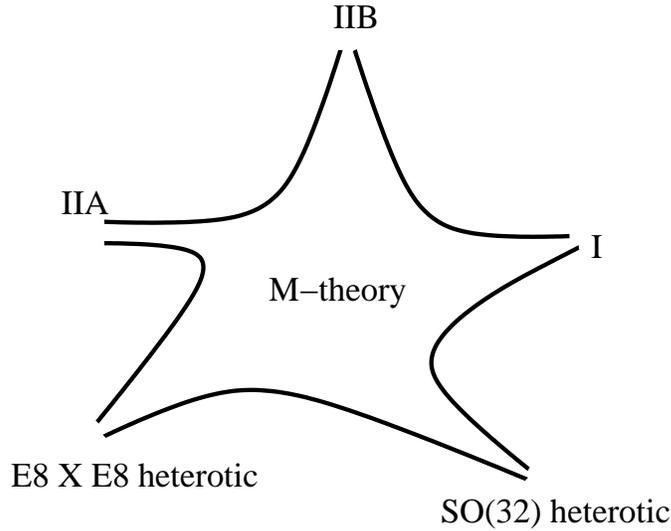}
\caption{Phases of M-theory.} 
\label{f9}
\end{center}
\end{figure}

Using various known dualities between different compactification
of different string theories one can now argue that all five
string theories are different ways of describing a single theory.
This theory has been given the name
M-theory. Different compactifications of
different string theories which are not related by duality are to be
regarded as different phases of M-theory, much in the same way that
water, ice and steam are to be regarded as different phases of a
single theory,  -- the theory of water molecules.\footnote{One difference
between these two cases is that while for water the three phases are
stable for different values of temperature, pressure etc., different
compactifications of string theory are all stable phases at zero
temperature.} A schematic (and much simplified)
picture of the phases of M-theory 
has been shown in Fig~\ref{f9}. 
A point in this diagram represents a phase of M-theory, and the
five holes represent the five weakly coupled 
string theories through which we may
try to get a view of the different phases of the theory. In principle
any point can be viewed as an appropriate 
`compactification'  of any of 
the five string theories, but clearly if we consider 
a point near one of the windows, -- representing the
corresponding string theory
with small value of the coupling constant, --
we have a better view of the point
from that window. Understanding what lies in the interior 
of the phase diagram, representing phases of M-theory which
cannot be viewed as weakly coupled theories from the viewpoint
of any of the five string theories, is one of the most challenging
problem for the present day string theorists.

Thus the problem of connecting M- theory to nature 
reduces to:
\begin{enumerate}
\item Demonstrating that there is a phase of M-theory that describes
exactly the nature that we observe.
\item Explaining why nature exists in this particular phase and not in any
other phase.
\end{enumerate}
Both issues are currently under active investigation by many researchers.
I shall end this talk by describing some speculative ideas on the 
second issue.

It has recently been found that M-theory 
has certain metastable phases.
These metastable phases are analogous to the supercooled or
superheated phases of matter. Consider for example the case of a
supercooled water, -- water below the normal freezing point.
As long as there is no disturbance
the water remains as water, but a small disturbance in any part
of the system will make a small region around that part condense
into the more stable ice phase. This small region of ice will then
expand inside the water and eventually convert 
the whole water 
into ice. Similarly the metastable phases of 
M-theory
have the property that occasionally some regions of the universe
in this phase may make transition into a more stable phase, and this
region then grows with time, converting the surrounding region into
the more stable phase. 

There is however a crucial difference between the way a 
metastable phase of M-theory behaves and a metastable phase
of a normal fluid behaves. The metastable phases of M-theory
which are relevant for our discussion 
have an additional 
property 
that
if any region of the universe 
is in that phase, it expands rapidly  as
a consequence of the laws of general theory of relativity.
In technical terms we say that these phases have positive
values of the cosmological constant, -- a constant that Einstein
had introduced into the equations for general relativity and later
abandoned due to lack of experimental evidence.\footnote{Recent
experiments have found that our universe has a small
but non-zero value of the cosmological constant.
Thus Einstein was right after all!
The phases of M-theory which we are
discussing here have much larger values of the cosmological
constant.}
Often the rate
of expansion of the universe due to this cosmological constant
term turns out to be much faster than the rate of expansion of the
bubbles of more stable phases which might form inside these
metastable phases.

\begin{figure}
\begin{center}
\leavevmode
\epsfysize 7cm
\epsfbox{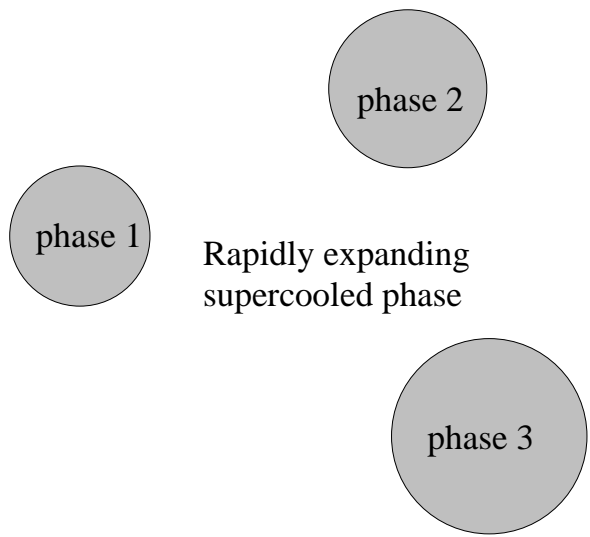}
\caption{State of the universe} 
\label{f10}
\end{center}
\end{figure}

Let us now combine these two facts about the metastable phases
of M-theory, and study how the universe will evolve if any region
of the universe happens to be in such a metastable phase of M-theory.
First of all, due to the cosmological constant term such a region of
the universe will expand very rapidly. At the same time in different
parts of the universe small regions of more stable phases will 
form\footnote{Even if there is no external disturbance, the laws of
quantum mechanics predict that there will be some intrinsic
disturbance in the universe which causes some randomly chosen
regions to form small bubbles of more stable phases.} which will
then grow, converting the surrounding region of the universe into
the more stable phase. In fact inside different bubbles 
we may have different stable
phases of M-theory. In a normal fluid this process will stop when
the walls of the expanding bubble eventually collide; and 
eventually the entire
fluid will be converted to
the most stable of all the phases.
However in the current situation this never happens since
the universe as a whole is expanding rapidly due to the
cosmological constant. Thus the process continues 
{\it ad infinitum}; the original universe keeps on expanding, and
more and more bubbles of stable phases form in different regions
of the universe. Eventually every possible phase of M-theory is
realized inside one or more bubbles. This situation has been depicted
in Fig.~\ref{f10}.

In this picture, no single phase of M-theory is prefered by nature.
The world that we see around us exists in a particular phase simply
because
we happen to live in this part of the world.
If we had lived in another part of the world we would see a different
phase.
Of course, in most of the phases of M-theory life as we know would be
impossible, and so nobody would be there to observe these phases.
But that is another matter!

\sectiono{Summary} \label{s6}

There are various  aspects of string theory which I have left
out of our discussion. These include string theory analysis
of black hole entropy, duality between string theory and gauge
theory etc.
My main focus in this article has been to 
explain how string theory brings us closer
to Einstein's dream.  However we are still quite far
from realizing our final goal of finding a complete theory of 
elementary constituents of matter. It is up to the present 
and the future
generation of string theorists to carry the theory forward towards this
goal. It will be an uphill task but worth the effort.

\end{document}